\documentstyle[12pt]{article}

\font\blackboard=msbm10 at 12pt
\font\blackboards=msbm7
\font\blackboardss=msbm5
\newfam\black
\textfont\black=\blackboard
\scriptfont\black=\blackboards
\scriptscriptfont\black=\blackboardss

\def\junk#1{{}}

\newcommand{\ba}{\begin{array}}
\newcommand{\ea}{\end{array}}
\newcommand{\be}{\begin{equation}}
\newcommand{\ee}{\end{equation}}
\newcommand{\bea}{\begin{eqnarray}}
\newcommand{\eea}{\end{eqnarray}}
\newcommand{\beas}{\begin{eqnarray*}}
\newcommand{\eeas}{\end{eqnarray*}}

\def\identity{{\rlap{1} \hskip 1.6pt \hbox{1}}}

\def\laplace{{\kern1pt\vbox{\hrule height 1.2pt\hbox{\vrule width 1.2pt\hskip
  3pt\vbox{\vskip 6pt}\hskip 3pt\vrule width 0.6pt}\hrule height 0.6pt}
  \kern1pt}}
\def\scriptlap{{\kern1pt\vbox{\hrule height 0.8pt\hbox{\vrule width 0.8pt
  \hskip2pt\vbox{\vskip 4pt}\hskip 2pt\vrule width 0.4pt}\hrule height 0.4pt}
  \kern1pt}}

\def\roughly#1{\raise.3ex\hbox{$#1$\kern-.75em\lower1ex\hbox{$\sim$}}}

\def\str{{\rm STr} \,}

\def\ht{{\hat{\cal T}}}

\def\st{{\tilde{\cal T}}}

\def\itt{{\cal  T}}

\newcommand{\NP}{{\em Nucl.\ Phys.\ }}
\newcommand{\PL}{{\em Phys.\ Lett.\ }}
\newcommand{\PR}{{\em Phys.\ Rev.\ }}

\newcommand{\PRL}{{\em Phys.\ Rev.\ Lett.\ }}

\newcommand{\gone}[1]{}
\begin{document}
\pagestyle{plain}
\setcounter{page}{1}

\baselineskip16pt

\begin{titlepage}

\begin{flushright}
PUPT-1799\\
hep-th/9806066
\end{flushright}
\vspace{20 mm}

\begin{center}
{\Large \bf Three-graviton scattering in Matrix theory revisited}

\vspace{8mm}

\end{center}

\vspace{3 mm}

\begin{center}

Washington Taylor IV\footnote{Address after August 1998: Center for
Theoretical Physics, Massachusetts Institute of Technology, Cambridge
MA 02139} and Mark Van Raamsdonk

\vspace{3mm}

{\small \sl Department of Physics} \\
{\small \sl Joseph Henry Laboratories} \\
{\small \sl Princeton University} \\
{\small \sl Princeton, New Jersey 08544, U.S.A.} \\
{\small \tt wati, mav @princeton.edu}

\end{center}

\vspace{9mm}

\begin{abstract}

We consider a subset of the terms in the effective potential
describing three-graviton interactions in Matrix theory and in
classical eleven-dimensional supergravity.  In agreement with the
results of Dine and Rajaraman, we find that these terms vanish in
Matrix theory.  We show that the absence of these terms is compatible
with the classical supergravity theory when the theory is compactified
in a lightlike direction, resolving an apparent discrepancy between
the two theories.  A brief discussion is given of how this calculation
might be generalized to compare the Matrix theory and supergravity
descriptions of an arbitrary 3-body system.

\end{abstract}

\vspace{18mm}
\begin{flushleft}
June 1998
\end{flushleft}
\end{titlepage}
\newpage

In the wake of the Matrix theory conjecture of Banks, Fischler,
Shenker and Susskind \cite{BFSS} and the AdS conjectures of Maldacena
\cite{Maldacena-AdS} there has been a remarkable body of evidence
found for a deep connection between supersymmetric Yang-Mills theories
and supergravity.  This connection can be made very precise for
certain terms arising in the perturbative one-loop expansion of Matrix
theory (See \cite{Banks-review,Susskind-review,WT-Trieste} for
reviews of Matrix theory).  Terms in the two-graviton Matrix theory
interaction potential of the form $v^4/r^7$ are protected by a
nonrenormalization theorem \cite{pss}.  These terms agree precisely
with the two-graviton interaction in 11-dimensional supergravity
\cite{DKPS,BFSS}.  This result can be made much more general: the
leading one-loop interaction potential between an arbitrary pair of
Matrix theory objects is of the form $F^4/r^7$, and it has been shown
that this matches precisely with the potential between an arbitrary
pair of objects in the linearized supergravity theory
\cite{Dan-Wati2}.  It seems likely that the full $F^4/r^7$ potential
is protected by supersymmetry, although this has not yet been
conclusively demonstrated.

At this point in time it is unclear how far 
the relationship extends between
perturbative Matrix theory and classical supergravity.  It has
been shown that finite $N$ Matrix theory corresponds to DLCQ M-theory
\cite{Susskind-DLCQ,Sen-DLCQ,Seiberg-DLCQ}.  This does not mean,
however, that finite $N$ Matrix theory necessarily agrees with DLCQ
supergravity \cite{Banks-review,Hellerman-Polchinski}.  If all terms
in the perturbative Matrix theory expansion are not protected by
nonrenormalization theorems, it seems necessary to understand the
large $N$ limit of the theory to verify the correspondence at higher
order.  It has been shown that certain higher-order terms in the
Matrix theory potential agree with supergravity.  Terms of the form
$F^4 X^n/r^{7 + n}$ precisely reproduce higher-moment interactions
from supergravity \cite{Mark-Wati,Dan-Wati2}.  Some two-loop terms of the form
$v^6/r^{14}$ seem to match with supergravity
\cite{bbpt}, although the situation regarding these terms is still
unclear (other terms of this form may break the correspondence
\cite{Vipul-rvu}; there cannot, however, be $v^6$ terms at higher
order in $1/r$ \cite{pss2}).  The first concrete suggestion of a
breakdown of the perturbative Matrix theory/supergravity
correspondence appeared in the work of Dine and Rajaraman
\cite{Dine-Rajaraman}.  These authors considered a 3-graviton
scattering process.  They found what appears to be a contradiction
between perturbative Matrix theory and classical supergravity; certain
terms which they expected in the classical supergravity theory could
not be reproduced at any order in Matrix theory.  In this letter we
address this problem; for certain terms considered by Dine and
Rajaraman we show that there is agreement between Matrix theory and
classical null-compactified supergravity.  The notation and
conventions used here roughly follow those of \cite{Dan-Wati2}.

As this letter was being written, we received the preprint \cite{ffi}
which addresses the same question.  The results of \cite{ffi} seem to
be in contradiction with those presented here, although both papers
reach the conclusion that Matrix theory and supergravity agree for
three-graviton scattering.  The results we describe here imply that the
terms in the diagram computed in \cite{ffi} will be canceled by terms
from other diagrams, so that there is no net contribution to the terms
of the form discussed in this letter.  This cancellation is
demonstrated explicitly in the Appendix.

The situation considered in \cite{Dine-Rajaraman}
is a scattering process with three incoming and three outgoing
gravitons.  Gravitons 2 and  3 are separated by a distance $r$ and
gravitons 1 and 3 are separated by a distance $R$ with $R \gg r$.  Dine
and Rajaraman argued that in supergravity there is a term in the
amplitude proportional to
\begin{equation}
\frac{ (v_1-v_2)^2 (v_2-v_3)^2 (v_1-v_3)^2}{R^7r^7} 
\label{eq:terms}
\end{equation}
where $v_i$ is the velocity of the $i$th graviton.  They showed that
no such term can appear in Matrix theory, and concluded that there
appears to be a discrepancy between Matrix theory and supergravity.
They suggested that the resolution of this discrepancy might reside in
the subtlety of the large $N$ limit.  In this letter, we argue that
the discrepancy may be eliminated by taking proper account of the
lightlike compactification of the supergravity theory.

In our discussion, we focus on the subset of the terms in
(\ref{eq:terms}) which are proportional to $v_1^4$.  We show that the
vanishing of these terms in Matrix theory is in fact in complete
agreement with what we would expect from the supergravity theory when
the theory is considered in the appropriate DLCQ context.  A complete
demonstration of the correspondence between the $1/r^7 R^7$ terms in
the perturbative supergravity potential and the Matrix theory
potential for the three-graviton system would involve a more
complicated calculation, and will be discussed in more detail
elsewhere.

We begin by explicitly showing that there are no terms proportional to
$v_1^4/R^7$ in the Matrix theory potential which depend on the
velocities of gravitons 2 and 3.  This can be demonstrated by first
integrating out the off-diagonal 1-2 and 1-3 fields, and then
integrating over the degrees of freedom in the 2-3 system.
Considering gravitons 2 and 3 as a single
system, which we denote by quantities with a hat, and graviton 1 as a second
system denoted by quantities with a tilde, the leading $1/R^7$ term in
the effective potential between the 2-3 system and graviton 1,
which arises from integrating out the 1-2 and 1-3 fields, is given by
the general two-body gravitational interaction formula
\cite{Dan-Wati2}
\begin{equation}
V_{{\rm one-loop}} =  - {15 R_{c}^2 \over 4 R^7} \left( \ht^{IJ} \st_{IJ}
- {1 \over 9} \ht^I{}_I \st^J{}_J\right) 
\label{eq:v-gravity} 
\end{equation}
where $\ht^{IJ}$  and $\st^{IJ}$ are defined in terms of the matrices
$\hat{X}^a, \tilde{X}^a$  describing the 2-3 system and particle 1
through the expressions
\begin{eqnarray*}
\itt^{--} & = & {1 \over R_{c}} \; \str  \left( \frac{1}{4} \dot{X}^a \dot{X}^a
\dot{X}^b \dot{X}^b+
{1 \over 4} \dot{X}^a \dot{X}^a F^{bc} F^{bc}  +
 \dot{X}^a \dot{X}^b F^{ac} F^{cb} \right. \nonumber\\ 
& & \qquad \qquad  \left.  +
{1 \over 4} F^{ab} F^{bc} F^{cd} F^{da}  -
\frac{1}{16} F^{ab} F^{ab}  F^{cd} F^{cd}  \right) \nonumber
  \\
\itt^{-a} & = & {1 \over R_{c}} \;\str \left(\frac{1}{2} \dot{X}^a \dot{X}^b \dot{X}^b +
{1 \over 4} \dot{X}^a F^{bc} F^{bc} 
                  + F^{ab} F^{bc} \dot{X}^c \right) \nonumber \\
\itt^{+-} & = & {1 \over R_{c}} \;\str \left(\frac{1}{2} \dot{X}^a \dot{X}^a + {1
\over 4} F^{ab} F^{ab} \right) \\ 
\itt^{ab} & = & {1 \over R_{c}}  \;\str \left( \dot{X}^a \dot{X}^b + F^{ac}
F^{cb} \right) \nonumber \\ 
\itt^{+a} & = & {1 \over R_{c}} \;\str \dot{X}^a \nonumber \\
\itt^{++} & = & {1 \over R_{c}} \;\str \identity =
{N \over R_{c}} \nonumber
\end{eqnarray*}
with $F_{ab}= -i[X_a, X_b]$.  (We use $I,J, \ldots$ to denote
11-dimensional indices and $a,b, \ldots$ to denote 9-dimensional
transverse indices.  $R_{c}$ is the compactification radius of the
$x^-$ direction.)

For graviton 1 we have
$\tilde{X}^a= (x_1^a+ v_1^a t)\identity_{N_1 \times N_1}$, so
\bea
 \st^{++} = {N_1 \over R_{c}} {\rm \hspace{0.55in}} \st^{+a}  &=   &
\frac{N_1}{R_{c}}  v_1^a
{\rm \hspace{0.55in}}  \st^{ab} = {N_1 \over R_{c}} v_1^a v_1^b 
\label{eq:stress-3}\\
 \st^{+-} = {N_1 \over 2R_{c}} v_1^2
\hspace{0.55in} \st^{-a}  &=  &
\frac{N_1}{2 R_{c}}  v_1^a v_1^2 
\hspace{0.45in}   \st^{--} = {N_1 \over 4R_{c}} v_1^4 \nonumber
\eea
If we restrict attention to the terms in the effective potential
which are proportional to $v_1^4/R^7$,  we are left with the
expression
\[
  - {15 R_c N_1 \over 16 R^7}  \langle \ht^{++} \rangle  v_1^4
\]
where by $\langle \cdot \rangle$ we mean the expectation value after
integration over the fluctuations of the 2-3 system.
Since $\ht^{+ +}= (N_2 + N_3)/R_{c}$ is a constant, however, we see that
there are no corrections at any order in $1/r$ to the one-loop term in
the interaction potential
\begin{equation}
 - {15 N_1 (N_2 + N_3) \over 16 R^7}  v_1^4 .
\label{eq:linear-potential}
\end{equation}
This agrees with the result of Dine and Rajaraman in
\cite{Dine-Rajaraman} that no terms of the form 
(\ref{eq:terms}) can be generated in Matrix theory after summing all
diagrams at any loop order.

We now discuss the corresponding terms in the classical supergravity
potential.  The leading $1/R^7$ term in the classical supergravity
potential between the  2-3 system and particle 1  arises from
diagrams in which a single graviton is exchanged.  Because the
propagator  for a graviton with  zero longitudinal momentum contains a
delta function $\delta (x^+ -y^+)$, the resulting interaction is a
classical instantaneous effective potential which takes precisely the
form (\ref{eq:v-gravity}), where $\ht^{IJ}$ and $\st^{IJ}$ are the integrated
components of the classical stress tensor for the 2-3 and 1 systems.
Note that the stress tensor $\ht$ in general 
contains terms proportional to $1/r^7$ which arise from 
internal graviton exchange processes between particles 2 and 3.
Such terms correspond in the classical theory to nonlinear corrections to
the stress tensor from the gravitational field itself.
Since  particle 1 is pointlike, its classical stress tensor
components are given by
\[
\st^{IJ} = \frac{p^I p^J}{p^+} .
\]
Using $p^+ = N_1/R_{c}$, $p^a = p^+  v_1^a$, and $p^- = p_a^2 / 2
p^+$ we arrive at the expressions (\ref{eq:stress-3}) for the
components of the integrated stress tensor of  particle 1.  To
determine the complete set of terms in the classical supergravity
potential which are proportional to $v_1^4/R^7$ it remains to evaluate
the component $\ht^{+ +}$ of the integrated stress-tensor for the 2-3
system.  This, however, is precisely the total conserved momentum of
the system in the  $x^-$ direction, which in the  null-compactified
supergravity theory corresponds to the quantity $(N_2 + N_3)/R_{c}$.
This shows that the only terms in the DLCQ supergravity three-graviton
potential proportional to $v_1^4/R^7$ are contained in the linearized
supergravity interaction (\ref{eq:linear-potential}), which is
precisely reproduced by Matrix theory.

We have therefore shown that Matrix theory and DLCQ supergravity are
in complete agreement for a subset of the terms in the three-graviton
potential.  This resolves, at least in part, the discrepancy found by
Dine and Rajaraman.  At this point, a few comments may be helpful in
clarifying the relationship between the analysis above and that in
\cite{Dine-Rajaraman}.  In \cite{Dine-Rajaraman}, the authors computed
a term proportional to  (\ref{eq:terms}).  This term
is also proportional to the product of the  momentum components in the
$x^-$ direction
\begin{equation}
k_1^+ k_2^+ k_3^+ \frac{ (v_1-v_2)^2 (v_2-v_3)^2 (v_1-v_3)^2}{R^7r^7} .
\label{eq:gravity-potential}
\end{equation}
In \cite{Dine-Rajaraman} there was an implicit assumption that $k_i^+
= N_i/R_{c}$ for $i = 1,2,3$.  However, this identification is not
necessarily correct.  Gravity is a nonlinear theory.  If we think of
supergravity as a field theory with a spin 2 field $h_{IJ}$ living in
a flat space-time background $\eta_{IJ}$, so that $g_{IJ}= \eta_{IJ}+
h_{IJ}$, then all components of the stress tensor, including the
momentum $\itt^{+ +}$, contain contributions nonlinear in the field
$h_{IJ}$.  The total momentum of the 2-3 system is then given by
$\ht^{+ +}=k_2^+ + k_3^+ +$ nonlinear terms.  These nonlinear terms
may be explicitly computed by expanding the stress tensor components
order by order in $h$ (see for example \cite{Weinberg}).  In the case
of interest here, it can be verified that there are nonlinear terms of
the form $k_2^+ k_3^+ (v_2-v_3)^2/r^7$ which contribute to the total
$\ht^{+ +}$.  There are numerous subtleties involved with defining
DLCQ theories (see for example \cite{Hellerman-Polchinski}).
Nonetheless, it seems reasonable to expect that the classical limit of
DLCQ supergravity will simply be the classical supergravity theory
compactified on a null circle.  In this classical theory, the
discretized quantity $\ht^{+ +}$ is the {\it total} momentum in the
compact direction, and includes all terms nonlinear in $h$.  It
follows that the nonlinear terms (\ref{eq:gravity-potential}) are
automatically included in the result (\ref{eq:linear-potential}) when
$(N_2+N_3)/R_c$ is identified with the total $\ht^{++}$.
This explains in detail how the apparent mystery of the missing terms
is resolved.

We conclude with a brief discussion of how  the analysis in this
letter might be generalized.  For an arbitrary 3-body system,  and for
the 3-graviton system in particular, there
are terms in the supergravity potential proportional to
\begin{equation}
\frac{\ht^{(2)}\st}{R^7 r^7} 
\label{eq:quadratic}
\end{equation}
where $\ht^{(2)}$ represents a contribution to the total stress tensor
of the 2-3 system arising from quadratic terms in the gravity theory.
For general objects in the supergravity theory there will also be
membrane current interactions arising from 3-form exchange.  In
general, we would expect to be able to reproduce terms of the form
(\ref{eq:quadratic}) by integrating out the 2-3 fields in the
expression (\ref{eq:v-gravity}).  In the case where we restrict
attention to terms proportional to the fourth power of the curvature
$F^4 (v^4)$ of  object 1, the integral over the 2-3 fields is
trivial because the integrand is a constant.  This is the case we have
considered in this letter.  For the other terms, the calculation is
more complicated.  A complete calculation of $\langle\ht^{IJ}\rangle$
would require, in particular, a generalization of the results of
\cite{Dan-Wati2}  describing the components of the stress tensor in the
presence of arbitrary background fermion fields.  Work in this
direction is in progress \cite{ktv}.

It was recently shown by Paban, Sethi and Stern \cite{pss2}
that the $v^6$ terms
in the two-graviton interaction potential are not renormalized.  This
gives some hope that the full collection of terms of the form
$F^6/r^{14}$, including those relevant for the general 3-body
calculation just mentioned, may similarly be protected by
supersymmetry.   If true, this would indicate that Matrix theory
correctly reproduces supergravity to quadratic order.  Recent work
\cite{deg} indicates that there may be problems with the velocity
expansion at order $v^8$ and beyond.  This suggests that the
correspondence with supergravity may break down at cubic order; more
detailed calculations are probably necessary, however, to be certain of this conclusion.

\vspace{0.2in}
\noindent
{\bf Note added:} Soon after the preprint of this paper appeared on
the net, a number of related papers appeared
\cite{Okawa-Yoneya,eg,msw,ffi2} discussing the three-graviton issue in
Matrix theory.  We have added this short note to clarify how this
paper is related to these other papers.  The paper \cite{eg} contains
an analysis of the cancellation of the diagram considered in
\cite{ffi}.  The result of \cite{eg} agrees with the result of the
Appendix of this letter, which states that in the background field
gauge we use here, this diagram is exactly canceled at one-loop order.
In \cite{ffi2} it is suggested that this particular diagram might not
cancel at one-loop order in a different gauge.  After summing all
diagrams, however, it is clear that physical quantities such as the
phase shift in the scattering amplitude should be gauge-invariant, so
one would expect terms of the form discussed in this paper to cancel
after summing all diagrams even in the gauge of \cite{ffi2}.  The
comprehensive analysis of Okawa and Yoneya \cite{Okawa-Yoneya} makes
it clear that there is an exact agreement between Matrix theory and
supergravity for the three-graviton scattering process (at least when
recoil is neglected).  In particular, by taking the limit of their
expression (4.4) in the case of three particles $A, B$ and $C$
where $r =x_{BC}\ll R =x_{AC}$ it can be
seen that no terms proportional to $v_{A}^4/r^7 R^7$ appear.  The
essential result of this letter is a simple argument for the vanishing
of this term both in Matrix theory and supergravity.  The vanishing of
this term in the supergravity theory indicates that the analysis of
Dine and Rajaraman in \cite{Dine-Rajaraman} was incomplete.  It should
be noted that the physical argument given in this letter for the
vanishing of this term in supergravity does not depend on the null
quantization of the gravity theory; in fact, the identification of
$(N_2 + N_3)/R_c$ with the integrated $T^{+ +}$ of the 2-3 system can
be seen simply from the fact that this is a conserved quantity in both
the Matrix and supergravity theories, which can be identified in this
way when particles 2 and 3 are widely separated, and which therefore
can also be identified as the particles come together.  Thus,
regardless of whether the supergravity is null-compactified or not,
the nonlinear terms we have discussed in $T^{++}$ (these terms are
also discussed in \cite{msw}) must automatically be included in $(N_2
+ N_3)/R_c$.  In summary, the results of this letter give a simple
physical argument indicating that the results of Dine and Rajaraman
are incomplete.  The result presented here agrees with the paper of
Okawa and Yoneya \cite{Okawa-Yoneya}, in which the question is settled
decisively by a careful calculation of the three-graviton scattering
amplitude in both theories.  The methods used in this paper can be
naturally extended to study interactions between $n$ arbitrary Matrix
theory objects, and to address the question of the Matrix
theory/supergravity correspondence at higher order.

\section*{Acknowledgments}

We would like to thank T.\ Banks, M.\ Dine, D.\ Kabat, P.\ Kraus, D.\ Lowe and
V.\ Periwal for helpful conversations and correspondence.  The work of
MVR is supported in part by the Natural Sciences and Engineering
Research Council of Canada (NSERC). The work of WT was supported in
part by the National Science Foundation (NSF) under contracts
PHY96-00258 and PHY94-07194.  WT would  like to thank the
organizers of the Duality 98 program and the ITP in Santa Barbara for
hospitality while this work was being completed.

\section*{Appendix}

In this appendix we describe explicitly the cancellation of the term
calculated in \cite{ffi}\footnote{After this work was completed we
were informed that this cancellation has been independently derived by
Echols and Gray \cite{eg}}.  In that paper a single diagram, the
``setting sun'' diagram, is computed.  This computation is performed in
two steps: first the very massive 1-2 and 1-3 fields are integrated
out, then the 2-3 fields are integrated out.  The cancellation of the
setting-sun diagram occurs already after integrating out the very
massive modes, when the corresponding setting-sun diagram is computed
with fermionic 1-2 and 1-3 fields.  This conclusion essentially
follows from the computation in \cite{Dan-Wati}, where the one-loop
calculation was performed for an arbitrary bosonic background.  The
connection between these calculations is somewhat subtle, however, as
they are performed using different propagators for the
off-diagonal fields.

In order to make the cancellation mechanism completely clear in the
case discussed in \cite{ffi}, we describe here explicitly the
diagrammatic cancellation using the same propagators as those used in
that paper.  We make the discussion slightly more general by
considering $n$ gravitons at positions $R_i$ where $R_1\gg R_i$ for $i
> 1$.  We assume that the $(N -1) \times (N -1)$ background matrices
describing gravitons $2-n$ contain arbitrary (but small) off-diagonal
terms, and write these matrices as
\[
X^a_{ij} = R_{ij}^a+K^a_{ij},\; \;\; \;\; \;1 < i,j \leq n
\]
where $R^a = {\rm Diag}(R^a_2, \ldots, R^a_n)$.  We wish to integrate
out the off-diagonal bosonic and fermionic fields $x_{i}^a$ ($1 \leq a
\leq 9$) and $\psi_{i}^\alpha$ ($1 \leq \alpha \leq 16$) (where we
have simplified notation by writing, for example, $x_i = x_{1i}$).
We ignore the gauge fluctuations in this calculation as they are not
relevant to  terms of the type computed in \cite{ffi}.
The relevant terms in the Lagrangian are
\begin{eqnarray*}
& &\frac{1}{R_c}  \left\{ 
\dot{x}^*_i \dot{x}_i
-x_i^* (r_i^2) x_i +
x_i^* (r_i \cdot K_{ij} + K_{ij} \cdot r_j -(K \cdot K)_{ij}) x_j
  -2(x^a_i)^*  ([Y_a,Y_b])_{ij} x_j^b\right. \\ 
& & \hspace{1in}\left.
+ i \psi_i \dot{\psi}_i
+\psi_i \gamma^a r^a_i \psi_i
-\psi_i \gamma^a (K^a)_{ij} \psi_j \right\}
\end{eqnarray*}
where we have defined
\[
(Y^a)_{ij} = (X^a)_{ij} - R^a_1 \delta_{ij}
\]
and
\[
r_i =  R_1-R_i,
\]
and we have suppressed some spatial and spinor indices $a$ and
$\alpha$.  Treating $r_i$ as the masses of the bosonic and fermion
fields, and the remaining terms as interactions, we can perform the
one-loop calculation order by order by summing over bosonic diagrams
with insertions of the quadratic vertices
of the forms $r \cdot K,K^2$ and $[Y,Y]$,
and fermionic diagrams with insertions of the quadratic vertex
of the form
$K \cdot \gamma$.  Terms with fewer than four
bosonic vertices of the $[Y,Y]$ type are all canceled, as 
shown in \cite{Dan-Wati}.  
In particular, the diagram calculated in \cite{ffi}
comes from a term of the form
\[
2\langle ((x^a_i)^* ([ Y^a, Y^b])_{ij} x_j^b) \cdot
((x^c_k)^* ([Y^c,Y^d])_{kl} x_l^d)
\rangle .
\]
Using the leading part of the bosonic propagator
\cite{Becker-Becker,ffi}
\[
\Delta ( t_1,t_2 |r^2) = \frac{1}{2r}  {\rm e}^{-r | t_1-t_2|}
\]
we find that
this term gives a contribution of
\begin{equation}
\frac{2}{r_i r_j (r_i + r_j)} 
\left\{ ([R^a,K^b])_{ij} 
 ([R^b,K^a])_{ji}  -
 ([R^a,K^b])_{ij} 
 ([R^a,K^b])_{ji} \right\}
\label{eq:term}
\end{equation}
as well as terms of the form $[K,K]^2$.  The expression (3.12) in
\cite{ffi} is a piece of the term (\ref{eq:term}) in the case $n = 2$.

Now consider the fermionic loop diagram with two insertions of
$K^a \gamma^a$.  The leading fermionic propagator contains two terms,
one containing a $\theta$ function and no $\gamma$ matrices, and the
other given by
\[
\frac{r^a \gamma^a}{2 r}  {\rm e}^{- r | t_1-t_2|} .
\]
Using only  this part of the propagator, we get a contribution
\[
-\frac{1}{4 r_i r_j (r_i + r_j)} 
{\rm Tr}\; (\gamma^a \gamma^b \gamma^c \gamma^d
r^a_i K^b_{ij} r^c_j K^d_{ji})
\]
The trace can be rewritten in the form
\begin{eqnarray*}
{\rm Tr}\; (\cdot) &  = &
16 \;{\rm Tr}\; \left( r^aK^br^bK^a-r^a K^br^a K^b + r^a K^ar^b K^b
\right) \\
&= & 8 \left(
([R^a,K^b])_{ij} 
 ([R^b,K^a])_{ji}  -
 ([R^a,K^b])_{ij} 
 ([R^a,K^b])_{ji}\right)  \\
& &
+ 8 \left( (r \cdot K + K \cdot r)_{ij}(r \cdot K + K \cdot r)_{ji}
\right)
-16 \;r_j^2  (K \cdot K)_{jj}
\end{eqnarray*}
The first term in this expression precisely cancels (\ref{eq:term}).
The remaining terms are higher moments of canceling diagrams, and  are
canceled by other one-loop diagrams.
The second term is canceled by the bosonic diagram with two $(r \cdot
K)$ insertions, and the third term is canceled by a combination of the
bosonic term with a single $K \cdot K$ term and the fermionic diagram
with two $K \cdot \gamma$ insertions and theta functions in the propagators.

We have thus shown  explicitly, using the same propagator structure as
in \cite{ffi}, that in general all the bosonic one-loop terms with two
$[Y,Y]$ insertions are canceled by a combination of bosonic and
fermionic diagrams.  In particular, this shows explicitly that the
term found in expression (3.12) of \cite{ffi} is canceled simply by
integrating out the very massive 1-2 and 1-3 modes, in accord
with the results of Dine and Rajaraman and the
discussion in the main text of this letter.

\bibliographystyle{plain}

\end{document}